\documentclass[useAMS,usenatbib,referee]{mn2e}
\usepackage{epsfig}
\usepackage{natbib}

\title[TRIMOR and its application to HD\,188753] {TRIMOR---three-dimensional correlation technique to
  analyze multi-order spectra of triple stellar systems; Application
  to HD\,188753}

\author[T. Mazeh et al.]
  {T. Mazeh$^{1,2}$,
  Y. Tsodikovich$^1$,
  Y. Segal$^{1,3}$,
  S. Zucker$^4$,
  A. Eggenberger$^{6}$,
\newauthor
  S. Udry$^5$ and
  M. Mayor$^5$\\
  $^1$ School of Physics and Astronomy, Raymond and Beverly Sackler
  Faculty of Exact Sciences, Tel Aviv University, Tel Aviv, Israel\\
  $^2 $Present temporary address: Radcliffe Institute for
Advanced Studies at Harvard \\ 
$^3$ Present address: Department of Applied Physics, Yale University,
New Haven, Connecticut 06520-8284 \\
  $^4$ Department of Geophysics and Planetary Sciences, Raymond and
  Beverly Sackler Faculty of Exact Sciences,\\
  \ \ Tel Aviv University, Tel Aviv, Israel\\
  $^5$ Observatoire de Gen\`eve, Universite de Geneve, 51 ch. des
  Maillettes, 1290 Sauverny, Switzerland\\
  $^6$ Laboratoire d'Astrophysique de Grenoble, 
       UMR 5571 CNRS/Universit\'e Joseph Fourier, BP 53, 38041 Grenoble Cedex
       9, France
  }
\date{Released 2008 Xxxxx XX}

\pagerange{\pageref{firstpage}--\pageref{lastpage}} \pubyear{2007}

\def\LaTeX{L\kern-.36em\raise.3ex\hbox{a}\kern-.15em
    T\kern-.1667em\lower.7ex\hbox{E}\kern-.125emX}

\begin{document}

\label{firstpage}

\maketitle

\begin{abstract}
This paper presents a new algorithm, TRIMOR, to analyse {\it
multi-order} spectra of triple systems. The algorithm is an extension
of TRICOR, the three-dimensional correlation technique that derives
the radial velocities of triple stellar systems from single-order
spectra. The combined correlation derived from many orders enables the
detection and the measurement of radial velocities of faint tertiary
companions. The paper applied TRIMOR to the already available spectra
of HD\,188753, a well known triple system, yielding the radial
velocities of the faintest star in the system. This rendered the close
pair of the triple system a double-lined spectroscopic binary, which
led to a precise mass-ratio and an estimate of its inclination. The
close-pair inclination is very close to the inclination of the wide
orbit, consistent with the assertion that this triple system has a
close to coplanar configuration.
\end{abstract}


%=====================
\section{Introduction}
%=====================

TODCOR (TwO-Dimensional CORrelation) is a two-dimensional correlation
technique to derive the radial velocities of the two components of
spectroscopic binaries \citep{zm94}. TODCOR was a natural extension of
the one-dimensional correlation technique \citep{griffin67,
simkin74,td79} which turned out to be very effective for deriving
radial velocities out of spectra with low signal-to-noise ratio
(SNR). TODCOR calculates the correlation of an observed spectrum
against a combination of two templates with different shifts,
resulting in a two-dimensional correlation function, whose peak
simultaneously identifies the radial velocities of both primary and
secondary components. Various studies have applied TODCOR successfully
to many binary spectra, yielding double-lined solutions
\citep[e.g.,][]{mazeh95, metcalfe96, torres97, zucker2004}, some of
them with quite faint companions \citep[e.g.,][]{mazeh97, prato02,
mazeh03}.

\citet{TRICOR} extended TODCOR to analyze spectra of triple-lined
systems by calculating the correlation of an observed spectrum against
a combination of {\it three} templates with three different
shifts. The peak of the resulting three-dimensional correlation
simultaneously identifies the radial velocities of the primary,
secondary and tertiary components. The new algorithm, TRICOR, was
successfully applied to the spectra of a few triple systems, yielding
triple-lined solutions \citep[e.g.,][]{torres95,
jha00}. \citet{torres07} extended the algorithm to analyse spectra of
quadruple systems.

The importance of TRICOR arises from the prevalence of short-period
binaries accompanied by a distant companion, as is becoming evident
from a few studies (e.g., \citealt{mm87}, and in particular
\citealt{tokovinin07}; see a review by \citet{mazeh08}). The
velocities of the three components contribute to a better
understanding of the orbital geometry of the triple system and the
dynamics of the three stars \citep[e.g.,][]{torres06}.

Both TODCOR and TRICOR aim to analyze spectra composed of one order
only. However, many of the modern spectrographs produce nowadays
multi-order spectra, some of which have as many as a few dozen orders
\citep[e.g.,][]{pepe00}. To optimize the use of multi-order
spectra for deriving accurate velocities and enabling us to detect faint
companions, we should include in the analysis all the spectral
information spread over the different orders. Therefore, \cite{TODMOR}
developed TODMOR, which can handle double-lined Multi-ORder spectra
for the analysis of double-lined systems. Using TODMOR, \citet{TODMOR,
zucker04} could discover an evidence for a planet and a brown dwarf
hidden in the spectra of the double-lined binary system HD\,41004.

The next step is to develop TRIMOR, a three-dimensional correlation
function to handle multi-order spectra.  This paper explains the
basics of TRIMOR, which is conceptually easy to construct, once we
have at hand TODMOR and TRICOR. In its second part, the paper applies
the algorithm to a set of spectra of HD\,188753, which is a well known
triple system \citep{Griffin77} that attracted new interest lately
\citep{konacki05}. The spectra analyzed here were obtained already by
\cite{Eg07} for a different study that focused on a careful analysis
of the radial velocity of the brightest star in the system. Applying
TRIMOR to those spectra enabled us to derive the velocities of the
faintest companion, rendering the close pair of the system a
double-lined binary.

Section 2 presents the principles of TRIMOR, Section 3 applies TRIMOR
to the already obtained spectra of HD\,188753, while Section 4 derives
the relative inclination of the system. Section 5 discusses briefly
the results obtained for HD\,188753 and future applications of TRIMOR.

%
%
%
%=========================
\section{TRIMOR Algorithm}
%=========================
%
\subsection{The three-dimensional correlation}
%----------------------------------------------

Assume $f(n)$ is an observed single-order spectrum of a triple system,
where $n$ is the wavelength. As $f(n)$ is composed of three spectra,
we wish to construct a model, $m(n)$, by calculating a spectrum
composed of $g_1$, $g_2$ and $g_3$ --- three templates for the
primary, secondary and tertiary components of the triple system,
respectively. Given the templates, the model
\begin{equation}
m(n)=g_1(n-s_1)+\alpha g_2(n-s_2)+\beta g_3(n-s_3)
\end{equation}
has five basic parameters: $s_1, s_2$ and $s_3$, the Doppler shifts of
the three templates, and $\alpha$ and $\beta$, the flux ratios of
$g_2$ and $g_3$ relative to $g_1$, respectively. The correlation
--- $f(n)\ast m(n)$, between the observed spectrum $f(n)$ and the model
$m(n)$, measures the goodness-of-fit of the model. Note that the
correlation is not sensitive to any multiplication factor of the model
function, and therefore only the {\it relative} flux of the three
templates is relevant to our analysis.

TRICOR considers the correlation between the observed spectrum and the
model as a function of all possible values
of $s_1, s_2$ and $s_3$, generating a three-dimensional correlation
function:

\begin{equation}
R^{(3)}(s_1,s_2,s_3;\alpha,\beta)=f\ast 
\left[ g_1(n-s_1)+\alpha
g_2(n-s_2)+\beta g_3(n-s_3) \right] \ ,
\end{equation}
where $\alpha$ and $\beta$ are considered external parameters that
have to be determined {\it a priori}. Alternatively, for each set of
values of $s_1, s_2$ and $s_3$ we can find analytically the values of
$\alpha$ and $\beta$ that maximize the correlation (See Appendix A in
\cite{TRICOR} for details).  The location of the maximum of
$R^{(3)}(s_1,s_2,s_3)$ identifies simultaneously the radial velocities
of the three components. To reduce the amount of computation we avoid
direct calculation of the three-dimensional correlation and instead
calculate only six one-dimensional correlations as detailed in
\cite{TRICOR}.

To consider multi-order spectra we combine the correlations of the
different orders according to the maximum-likelihood approach
developed by \cite{Zucker03}:
\begin{equation}
R^{2}(s_1,s_2,s_3)=1-\lbrace\prod^M_{i=1}[1-R_i^2(s_1,s_2,s_3)]\rbrace^{1/M}
\ ,
\end{equation}
where $R_i$ is the three-dimensional correlation function of the
$i^{th}$ order and $M$ is the number of orders.

When considering multi-order correlation functions, the values of
$\alpha$ and $\beta$ vary from order to order, and therefore, $\alpha$
and $\beta$ are in fact vectors of values, $\{\alpha_i, i=1,M\}$ and
$\{\beta_i, i=1,M\}$. Finding numerically different $\alpha$ and
$\beta$ values for each order would greatly increase the number of
fitted parameters. We, however, can reduce drastically the complexity
of the calculation by taking advantage of the fact that for given two
stars with known spectral types the flux ratio at a given wavelength
is known approximately \citep[e.g.,][]{Pickles98}. We therefore
pre-calculated tables containing the flux ratio at each order for many
possible spectral-type pairs in our template library, using
\citet{Pickles98} tables. In other words, the choice of templates
determined the values of $\{\alpha_i; i=1,M\}$ and $\{\beta_i;
i=1,M\}$ in our analysis.

\subsection{Choosing the parameters of the correlation function}

To derive the most accurate velocities out of a given set of observed
spectra we have to carefully choose the parameters of the model
function. We first search for the best templates to fit the observed
spectra. This step depends on the available library of templates that
fit the spectral range and resolution of the observed spectra. We
first find the template for the primary, as this template has the
largest contribution to the model function. We find the best template
by running a one-dimensional correlation with each template candidate
against all the observed spectra and choose the one that yields the
highest correlation peak, averaged over all spectra. We then find the
template for the secondary by running TODMOR with all template
candidates over all observed spectra, and only at the third stage we
choose the template for the tertiary. In many cases the contribution
of the tertiary to the correlation function is minimal and we have to
choose the template that leads to the best orbital solution of the
tertiary, usually judged by the size of its residual RMS.

As explained above, the choice of the three templates determines the
flux ratio between the templates that we use. However, to maximize the correlation
we allow a multiplicative factors $f_{\alpha}$ and $f_{\beta}$ of the
order of unity which we apply to the pre-determined flux ratios of all
orders. The factors are determined numerically to maximize the average
of the correlation peak.

In some cases the actual stars in the system have rotational
broadening larger than their templates. To correct for that we try to
insert numerically some {\it additional} rotational broadening to the
three templates.  This is done by convolving the template spectra with
a broadening function, based on the Fourier transformation of
\citet{sg76}. Here again we first determine the rotational broadening
of the primary template, then the secondary and at last the
tertiary. The values of the rotation of the three templates are chosen
to maximize the averaged correlation peak.

In finding the best parameter values of the model function, we always
choose the ones that maximize the correlation peak {\it averaged over
all observed spectra}.  Because the values of all these parameters are
based on the whole set of observed spectra we consider these
parameters as global ones. In some cases, however, we prefer to
perform the averaging only over the high-SNR spectra, as the low-SNR
spectra add noise to the process.

To conclude, we fit eight global parameters to the whole set of observed
spectra: the best three spectral templates, two multiplicative factors
to be applied to the flux tables, and three rotational broadening
velocities. After finding these eight global parameters, we consider
the three-dimensional correlation for each observed spectrum, find its
maximum and derive the three velocities of the three stars for each
spectrum. 

The uncertainties of the three velocities were derived as explained by
\citet{Zucker03}, by deriving the inverted Hessian of the likelihood
of the velocities. However, these errors were proven to be
overestimated in quite a few cases \citep[e.g.,][]{Eg07}, and we had to
rescale them by using the RMS of the residuals of the orbital
solution. In this case, we kept only the relative sense of the derived
error estimate, which probably depended mainly on the SNR of the
obtained spectra.

%
%
%
%========================================================
\section{Application of TRIMOR to the case of HD\,188753}
%========================================================
%
%
%
\subsection{The system HD\,188753}
%----------------------------------
%
HD\,188753 (HIP 98001) is a bright triple system, $7.4$ V-mag, at a
distance of $46$ pc \citep{soderhjelm99}. \cite{Hough99} was the first
to discover that the system was a visual binary, with a separation
between the two components, A and B, of $\sim0.3$ arc-sec.  The AB
pair is characterized by a period of $25.7$ years, a semi-major axis
of $12.3$ AU and an eccentricity of $0.5$ \citep[][see also
\citealt{konacki05}]{soderhjelm99}. Later \cite{Griffin77} discovered that
the system contained a spectroscopic binary with an orbital period of
$155$ days and an eccentricity of $0.26$.

Recently, HD\,188753 attracted interest when \cite{konacki05} reported
a radial-velocity modulation of the primary --- A, with a period of
$3.35$ days and an amplitude that implied an unseen companion with a
minimum mass of $1.14$ Jupiter mass. The detection of a close-in
Jupiter-like planet in the triple system HD\,188753 posed a problem
for the widely accepted planet formation theory (e.g.,
\citealt{pfahl05,jc07}). However, the existence of the close-in planet
is now a matter of debate, after \cite{Eg07} obtained their own ELODIE
spectra, the analysis of which showed no evidence to support the
conjecture of a planet in the system.

\cite{Eg07} used TODMOR to derive the velocities of
A, the bright distant star, and Ba, the brighter star of the close
pair. The main goal of \cite{Eg07} was to study the radial velocity of
component A, and to confirm or refute the planet
conjecture. Nevertheless, their spectra included information on the
close pair, which before this work remained a single-lined
spectroscopic binary. To demonstrate the capability of TRIMOR we
re-analysed here the spectra obtained by \cite{Eg07} and derived the
velocities of A, Ba and Bb, the latter being the faintest component of
the triple system, rendering the close binary a double-lined
spectroscopic system.

\subsection{The TRIMOR radial velocities}
%----------------------------------------
%
The data we analyzed here comprised the ELODIE spectra of HD\,188753,
obtained by \citet{Eg07} between July 2005 and March 2006.  ELODIE is
a fiber-fed spectrograph at the Observatoire de Haute-Provence
(France) with a fiber of $2''$ in diameter \citep{Baranne96}. With a
resolution of $\lambda/\Delta\lambda = 42\,000$ it covers the
wavelength range 3850--6800\,\AA\, with 67 orders.  The analysis used
32 ELODIE orders within the spectral range 4810--6800\,\AA, after
having excluded orders that are heavily polluted by telluric lines.
We also excluded the blue orders because in these orders we expected
the secondary and tertiary signals to be too weak relative to the
signal of the hotter and bluer primary.

Our library of templates included high-SNR ELODIE and CORALIE spectra,
with spectral types that run from F8 to M6.  To choose the best
configuration of templates and rotational broadening we proceeded as
detailed above. In our case, the tertiary template was so faint that
its impact on the correlation peaks was minimal. Therefore, we carried
out the analysis with the different possible tertiary templates
through the orbital solution, and chose the one that minimized the
residual RMS of the tertiary orbit. 

We finally chose a G6 dwarf (HD\,224752) for A, a
K0V dwarf (HD\,225208) for Ba, and a K7 dwarf (GL\,338B) for Bb.  As
we do not have a table with flux ratios for every pair of spectral
types, we used for $\alpha$ our G5K0 table, for which the flux ratio
at 6800\,\AA\ was 0.53, and a multiplication factor, $f_{\alpha}$, of
$1.12$. For $\beta$ we used the G8M0 table, for which the flux ratio
at 6800\,\AA\ was 0.06, and a multiplication factor, $f_{\beta}$, of
$0.91$. In our model the fluxes of the primary, A, and the secondary,
Ba, are comparable, while the tertiary, Bb, is much fainter than the
other two. The {\it additional} rotational broadening velocities
chosen for the templates used were $0.0$ km\,s$^{-1}$, $2.0$
km\,s$^{-1}$ and $2.5$ km\,s$^{-1}$ for A, Ba and Bb, respectively.

As in \citet{Eg07}, we also noticed that when two of the three stars
had similar velocities, TRIMOR sometimes picked the wrong peak of the
correlation, yielding erroneous velocities. Three additional
spectra that were analysed by \citet{Eg07} yielded ambiguous
results.  We preferred not to include these velocities in our
analysis.  Thus we ended up using only $35$ spectra, with a typical
SNR of $55$ per pixel at 5500 \AA.

Our radial velocities and their error estimate for the three
components of HD\,188753 are given in Table~\ref{vels_table}. The
original error estimates for the velocities of Ba and Bb turned out to
be larger than the actual residuals of the orbital solution. This
effect appeared in a few other stellar systems that we have analysed
(see also \citet{Eg07} for similar effect with TODMOR). We therefore
adjusted the error estimates by multiplying the derived errors of Ba
by 0.34 and those of Bb by 0.26, so the normalized $\chi^2$ value of
the solution is unity for each of the three stars.

\begin{table}
\caption{The derived radial velocities for HD\,188753}
\begin{center}
\begin{tabular}{llll}
\hline\hline
JD-2450000 & Primary [$km\,s^{-1}$] & Secondary [$km\,s^{-1}$] & Tertiary [$km\,s^{-1}$] \\
\hline

$3575.472$ & $-22.826 \pm 0.094$ & $-36.766 \pm 0.065$ & $-2.44 \pm 0.41$ \\ 
$3576.470$ & $-22.811 \pm 0.088$ & $-36.632 \pm 0.061$ & $-2.66 \pm 0.39$ \\ 
$3577.478$ & $-22.818 \pm 0.096$ & $-36.605 \pm 0.065$ & $-2.16 \pm 0.42$ \\ 
$3585.448$ & $-22.826 \pm 0.105$ & $-34.441 \pm 0.079$ & $-4.76 \pm 0.47$ \\ 
$3586.457$ & $-22.874 \pm 0.118$ & $-34.079 \pm 0.086$ & $-5.84 \pm 0.56$ \\ 
$3587.405$ & $-22.870 \pm 0.104$ & $-33.657 \pm 0.076$ & $-5.98 \pm 0.49$ \\ 
$3588.433$ & $-22.953 \pm 0.088$ & $-33.341 \pm 0.059$ & $-7.10 \pm 0.43$ \\ 
$3589.484$ & $-22.916 \pm 0.119$ & $-33.026 \pm 0.082$ & $-6.68 \pm 0.54$ \\ 
$3590.467$ & $-22.864 \pm 0.108$ & $-32.461 \pm 0.077$ & $-7.74 \pm 0.48$ \\ 
$3591.505$ & $-22.866 \pm 0.101$ & $-32.078 \pm 0.070$ & $-8.27 \pm 0.45$ \\ 
$3594.487$ & $-22.861 \pm 0.097$ & $-30.890 \pm 0.067$ & $-10.27 \pm 0.43$ \\ 
$3595.420$ & $-22.801 \pm 0.092$ & $-30.462 \pm 0.065$ & $-10.41 \pm 0.43$ \\ 
$3596.448$ & $-22.794 \pm 0.095$ & $-29.985 \pm 0.066$ & $-10.73 \pm 0.47$ \\ 
$3667.380$ & $-22.782 \pm 0.091$ & $-10.066 \pm 0.066$ & $-36.27 \pm 0.47$ \\ 
$3669.299$ & $-22.777 \pm 0.080$ & $-10.168 \pm 0.058$ & $-35.60 \pm 0.44$ \\ 
$3686.258$ & $-22.823 \pm 0.082$ & $-14.845 \pm 0.057$ & $-30.57 \pm 0.36$ \\ 
$3690.264$ & $-22.971 \pm 0.085$ & $-16.871 \pm 0.062$ & $-28.18 \pm 0.45$ \\ 
$3693.256$ & $-23.008 \pm 0.095$ & $-18.548 \pm 0.068$ & $-26.47 \pm 0.47$ \\ 
$3694.234$ & $-23.021 \pm 0.096$ & $-19.150 \pm 0.066$ & $-25.62 \pm 0.44$ \\ 
$3711.296$ & $-22.977 \pm 0.107$ & $-31.103 \pm 0.076$ & $-9.27 \pm 0.56$ \\ 
$3714.245$ & $-23.142 \pm 0.123$ & $-33.111 \pm 0.087$ & $-7.43 \pm 0.64$ \\ 
$3715.215$ & $-23.108 \pm 0.093$ & $-33.523 \pm 0.065$ & $-6.50 \pm 0.46$ \\ 
$3719.212$ & $-23.107 \pm 0.106$ & $-35.348 \pm 0.078$ & $-3.58 \pm 0.52$ \\ 
$3723.247$ & $-23.029 \pm 0.087$ & $-36.272 \pm 0.063$ & $-2.20 \pm 0.41$ \\ 
$3724.234$ & $-23.026 \pm 0.105$ & $-36.443 \pm 0.074$ & $-2.02 \pm 0.48$ \\ 
$3728.231$ & $-23.039 \pm 0.084$ & $-36.768 \pm 0.058$ & $-1.75 \pm 0.38$ \\ 
$3807.689$ & $-22.980 \pm 0.123$ & $-10.297 \pm 0.087$ & $-35.43 \pm 0.59$ \\ 
$3809.682$ & $-22.954 \pm 0.123$ & $-10.015 \pm 0.082$ & $-35.93 \pm 0.58$ \\ 
$3810.675$ & $-23.002 \pm 0.115$ & $-9.921 \pm 0.077$ & $-36.45 \pm 0.55$ \\ 
$3871.585$ & $-23.375 \pm 0.091$ & $-34.266 \pm 0.062$ & $-4.46 \pm 0.39$ \\ 
$3873.594$ & $-23.373 \pm 0.125$ & $-35.187 \pm 0.082$ & $-2.95 \pm 0.53$ \\ 
$3895.597$ & $-23.438 \pm 0.097$ & $-33.774 \pm 0.067$ & $-5.04 \pm 0.44$ \\ 
$3896.581$ & $-23.439 \pm 0.091$ & $-33.390 \pm 0.063$ & $-5.65 \pm 0.40$ \\ 
$3899.588$ & $-23.363 \pm 0.088$ & $-32.074 \pm 0.061$ & $-7.08 \pm 0.39$ \\ 
$3900.559$ & $-23.423 \pm 0.133$ & $-31.911 \pm 0.089$ & $-7.16 \pm 0.68$ \\ 
\hline
\end{tabular}
\end{center}
\label{vels_table}
\end{table}

\subsection{The orbital solution}
%--------------------------------
%

To derive the orbital solution we solved for double-lined solution of
Ba and Bb, {\it simultaneously} with a linear drift of their
center-of-mass velocity. This added one more linear parameter to the
function fitted to the radial velocity measurements. For the A
velocities, we solved independently for a linear drift only. 

\begin{figure}
\centering
\epsfig{file=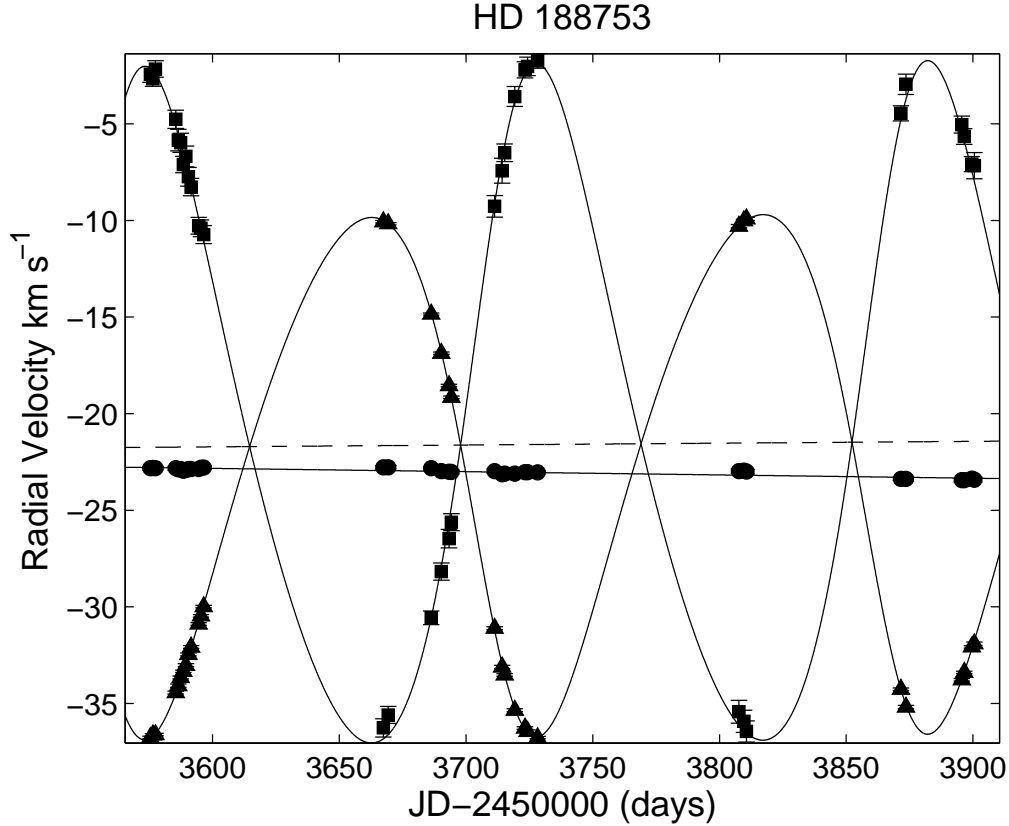,width=15cm}
\caption{Orbital solution for HD\,188753. The circles refer to A,
triangles to Ba and squares to Bb. Error bars in A are smaller than
the size of the data points. The solid straight line represents the
$25.7$-year orbital motion of A and the dashed line that of B. }
\label{vels}
\end{figure}

The orbital
solution based on the TRIMOR velocities is given in Table~\ref{tab1}
and plotted in Figure~\ref{vels}. The linear drift for A and B could
barely be detected, because of the relatively short time span of the
observations. All our derived values are consistent with the results
of \citet{Eg07}. Our error estimates of these values are smaller than
those of \citet{Eg07}, indicating another possible advantage of
TRIMOR.

%
%----------------------------------------------------------------
%Figure 1
%=========
%----------------------------------------------------------------

\begin{figure}
\centering
\epsfig{file=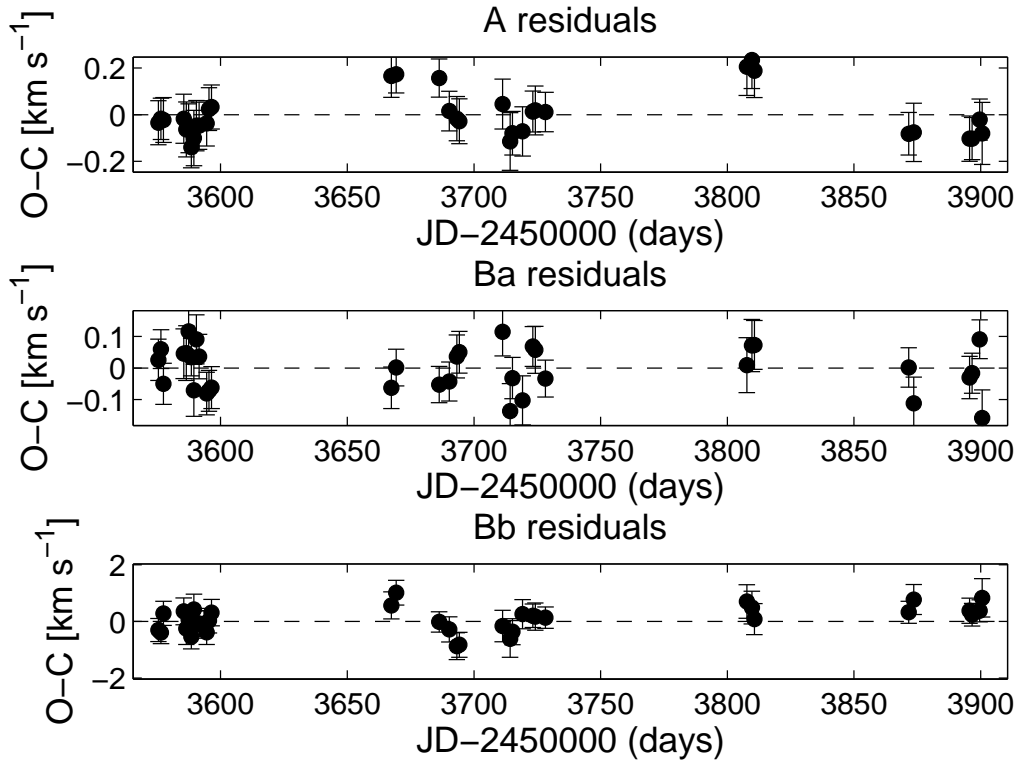,width=15cm}
\caption{The residuals of the orbital solution in Fig.~\ref{vels}. The
RMS for A is $98$ m\,s$^{-1}$, for Ba is $71$ m\,s$^{-1}$ and for
Bb is $0.5$ km\,s$^{-1}$.}
\label{res}
\end{figure}

The residuals of the three sets of velocities are plotted in
Figure~\ref{res}.  While the residual RMS of A and Ba is at
the level of $~0.1$ km s$^{-1}$, the RMS of the Bb residuals is of the
order of $0.5$ km s$^{-1}$. This is consistent with the relative
brightness of the three stars.

%----------------------------------------------------------------
%Figure2
%==========
%----------------------------------------------------------------

%----------------------------------------------------------------
\begin{table}
\caption {Orbital parameters for the 154-day spectroscopic orbit of 
HD\,188753\,B. The long-period orbital motion of the AB pair was taken into 
account by including a linear drift.}
\begin{center}
\begin{tabular}{llc}
\hline\hline
Parameter & Units & Value \\
\hline
$P$                         & (days)            & $154.448\pm0.015$\\
$T$                         & (JD-2400000)      & $53405.007\pm0.032$\\
$e$                         &                   & $0.175\pm0.001$\\
$\gamma$                    & (km\,s$^{-1}$)     & $-21.623\pm0.014$\\
$\omega$                    & (deg)             & $134.87\pm0.02$\\
$K_1$                       & (km\,s$^{-1}$)     & $13.479\pm0.017$\\
$K_2$			    & (km\,s$^{-1}$)     & $17.63\pm0.09$\\
Linear drift A               & (km\,s$^{-1}$\,yr$^{-1}$)  & $0.619\pm0.056$ \\
Linear drift B               & (km\,s$^{-1}$\,yr$^{-1}$)  & $0.346\pm0.039$ \\
\cline{1-3}
$N_{\rm meas}$              &                   & 35\\
rms (A)                         & (m\,s$^{-1}$)     & 98.3\\
rms (Ba)                         & (m\,s$^{-1}$)     & 72.1\\
rms (Bb)                         & (m\,s$^{-1}$)     & 455.6\\
\hline
\end{tabular}
\end{center}
\label{tab1}
\end{table}
%-------------------------------------------------------------------------

The new double-lined solution of the B system yields directly its mass
ratio: 
\begin{equation}
q_B=\frac{m_{Bb}}{m_{Ba}}=0.768\pm0.004 \ .
\end{equation}
The mass ratio of the wide orbit is:

\begin{equation}
q_{AB}=\frac{m_B}{m_A}=1.79\pm0.26 \ .
\end{equation}
The latter has a large error, because of the short time span of the
observations, but still is in agreement with \cite{Eg07}, who derived
the linear drift of $A$ with a different way.

\subsection{Search for evidence of a planet}
%--------------------------------------------

To search for an evidence of the planet suggested by \cite{konacki05},
we performed a search for periodicity in our derived radial velocities
of HD\,188753\,A, as did \cite{Eg07}.  In fact, the residuals of A
were slightly larger than those of B, and therefore could hide
additional real variability. To search for {\it any} periodicity we
derived a double-harmonic periodogram of the residuals
\citep{shporer07}, by fitting for each trial period a double-harmonic
function. The periodogram value for each trial period was taken as the
power (sum of the squared coefficients of the two harmonics) divided
by the $\chi^2$ goodness-of-fit parameter of the fitted light curve.
We preferred this analysis to the frequently used Lomb-Scargle test
\citep[e.g.,][]{scargle82}, as the radial velocity of the presumed
additional planet might not have a perfect circular orbit, and
therefore might have a periodic modulation with a few harmonics. Our
method is a further generalization of \citet{kurster09} method.

To estimate the statistical significance of a peak with an height $S$
we analysed 10,000 randomized set of residuals and found the fraction
of periodograms that had a peak higher than $S$. The randomized
residuals were derived by taking the actual residuals, with the actual
timings randomly permutated. Assuming a negligble red noise in the
data \citep[e.g.,][]{pont06}, this test gives the statistical
significance of any detection.

\begin{figure}
\centering
\epsfig{file=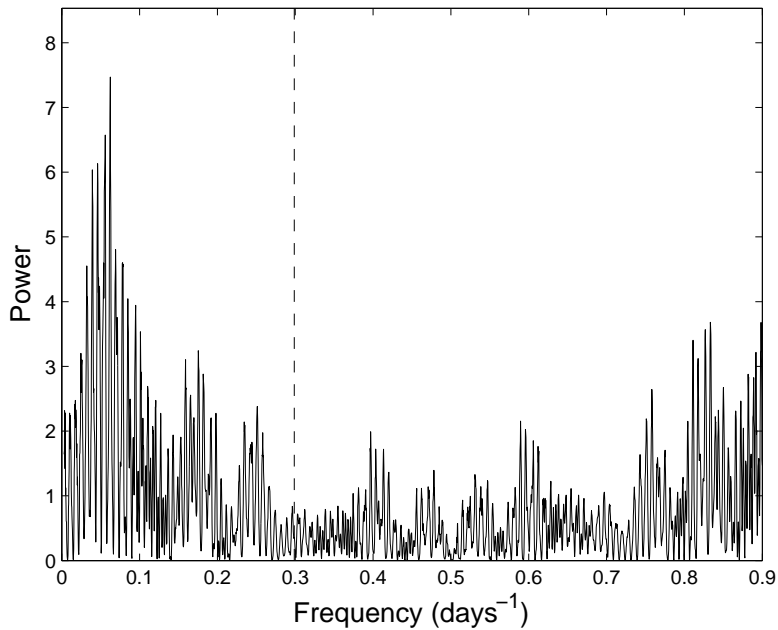,width=12cm}
\caption{Double-harmonic periodogram of the residuals around the linear
trend for HD\,188753 A. For this data set, the $1\%$ false alarm
probability corresponds to a power of $8.58$ and is represented by the
top of the box. The dashed line denotes the frequency of the planetary
signal reported by Konacki.}
\label{A_ps}
\end{figure}

The computed periodogram of the residuals around the linear trend for
component A are plotted in Figure~\ref{A_ps}. Our velocities for
HD\,188753\,A show no sign of a short-period signal with the period of
$\sim 3$ days, although some variability with a period of about 20
days might be present in the data. However, the periodagram peak might
reflect the window function of the data, and in any case is not
significant enough to draw any conclusion without further
observations.

%-----------------------------------------------------------------
%------------------------------------------------------------------

%
%
%
%===================================================
\section{The relative inclination of the two orbits}
%===================================================
%
%
%
HD\, 188753 is a special case, for which accurate astrometric,
photometric and spectroscopic information is available. We wish to use
this information and estimate the degree of alignment between the
orbital planes of the wide (astrometric) and the close
(spectroscopic) orbits. \citet{konacki05} shortly addressed this issue,
but he did not have yet a direct measurement of the mass ratio of the
close pair.

\cite{soderhjelm99} estimated the inclination of the wide orbit at
 about $34^{\circ}$. We need now to
derive the inclination of the spectroscopic binary so we can estimate
the relative inclination.

For any double-lined spectroscopic binary, the orbital inclination can
be written as:

\begin{equation}
 \sin i =  4.7 \times 10^{-3} \ 
\sqrt{1-e^2}
\left(\frac{K_1+K_2}{1\,\mathrm{km\,s}^{-1}}\right)
\left(\frac{P}{1\,\mathrm{day}}\right)^{1/3}
\left(\frac{M_1+M_2}{M_{\sun} }\right)^{-1/3}
\ .
\end{equation}
In this formula, only the total mass of the spectroscopic binary is
not known from the radial-velocity data. Deriving the total mass of
the binary from information not included in the orbital solution leads
to an estimate of the binary inclination.

We will use here three approaches to estimate the B-component
mass. As in many SB2s, we can derive the total mass, or at least
estimate it, through analyzing the spectra of the two stars.  In our
case we can take advantage of two additional known features of
HD\,188753. First, the binary is a component of a wide orbit with
known relative astrometric orbit. Therefore, the total mass of the
triple system is known. To derive the mass of the close pair we only
need the mass ratio of the wide orbit, which we can obtain from the
spectroscopic data, provided their time span is long enough
\cite[e.g.,][]{Eg07}.  Second, the parallax of the system is known
\citep{soderhjelm99}, because A is bright enough to be measured by
Hipparcos. Using the parallax, we can derive the expected magnitude of
HD188753Ba {\it and} HD188753Bb for each assumed mass of the Ba
component, using the mass ratio, $q_B$, obtained from the
radial-velocity solution.  We can then compare the expected magnitude
of the combined light of Ba and Bb with the observed $J$, $H$ and
$K_s$ 2MASS magnitudes of B obtained by \citet{konacki05}. In this way
we can find the actual mass of Ba, the single remaining independent
parameter in this exercise.

\begin{figure}
\centering
\epsfig{file=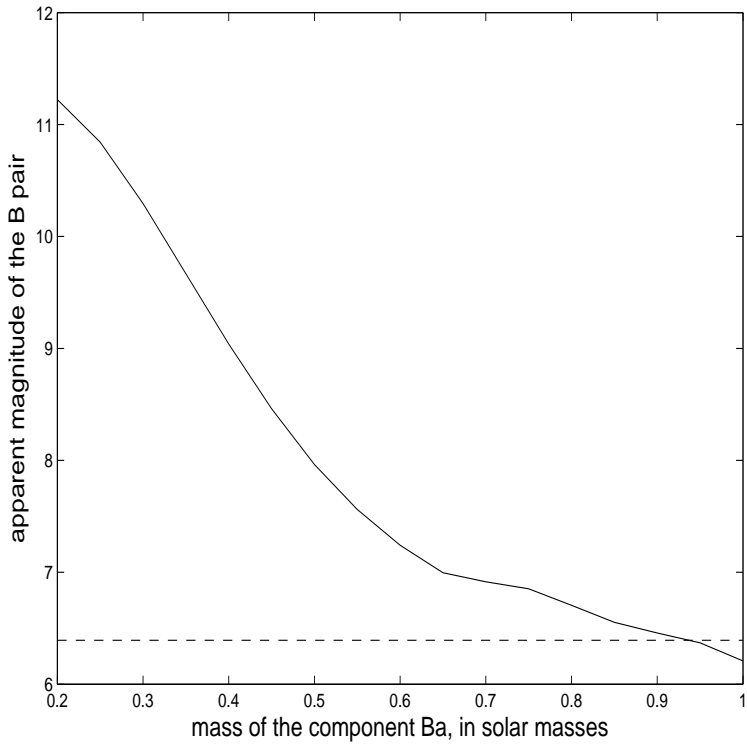,width=10cm,height=10cm}
\caption{For each assumed mass of the Ba component, the plot shows the
derived $K_s$ magnitude (in the 2MASS $K_s$ band), assuming the known
mass ratio $q_B$, and the known distance modulus of $3.31$. The dashed
line represents the observed $K_s$ magnitude.}
\label{Kmass}
\end{figure}

Table~\ref{sini_tab} presents the results of the three approaches.
For the spectral analysis approach we utilized the spectral types we
 used in our TRIMOR analysis (K0V and K7V) and 
a standard calibration \citep{hh81} to convert them to masses.  For the
astrometric approach we used the total mass of the system, $2.73
M_{\odot}$ \citep{soderhjelm99}, and the mass ratio derived by
\cite{Eg07}. For the photometric approach we used a parallax of
$\pi=21.9\pm0.6$ mas, the photometric calibration of \citet{bb88} and the
2MASS correction of \citet{carpenter01}. Figure~\ref{Kmass}
demonstrates this procedure for the $K_s$ band.  Error estimates are
not readily available for the photometric calibrations, but we can use
the variance among the results from the three photometric bands as a
proxy to the error.

%------------------------------------------------------------------
\begin{table}
\caption {The total mass of the close binary and its inclination,
based on three approaches}
\begin{center}
\begin{tabular}{lccc}
\hline\hline
approach & $M_{Ba}+M_{Bb}$ & $\sin i$ & $i$ \\
         &    [$M_{\sun}$] &          &   \\
\hline
spectral types obtained by TRIMOR & $1.40 \pm 0.15$ & $0.690 \pm 0.026$ & $43\fdg6 \pm 2\fdg1$ \\
astrometry of the visual pair     & $1.78 \pm 0.31$ & $0.637 \pm 0.038$ & $39\fdg6 \pm 2\fdg8$ \\
2MASS photometry                  & $1.634 \pm 0.019$ & $0.6557 \pm 0.0079$ & $40\fdg97 \pm 0\fdg60$ \\
\hline
\end{tabular}
\end{center}
\label{sini_tab}
\end{table}
%----------------------------------------------------------------------------

The results of the three approaches show a remarkable consistency,
indicating that the inclination of the close pair is around
$41\degr$. It is not easy to estimate the uncertainty of the derived
inclination, as the biases of the approaches are not well
known. Arbitrarily, we assign an error of $3\degr$ to our estimate
of the binary inclination.

The close-pair inclination found here, $41\degr\pm3\degr$, is close to
$34\degr$, the value given by \citep{soderhjelm99} for the wide orbit
inclination. This means that the minimal relative inclination (MRI)
between the two orbits is around $7\degr$. \citet{soderhjelm99} values
are not accompanied by error estimates, and therefore it is difficult
to assign an uncertainty to the MRI value of the triple
system. However, if we arbitrarily assign an error of $3\degr$ to the
wide orbit inclination, we can deduce an error of $\sim 5\degr$ to our
MRI. Our findings are therefore consistent with the MRI being close to
zero, and therefore the system might have coplanar configuration.

%
%
%
%====================
\section{Discussion}
%====================
%
%
%

In this paper we laid out the principles of TRIMOR, an algorithm to analyse
multi-order spectra of triple systems and deduce the radial velocities
of the three components. A few improvements of the algorithm could
still be applied, one of which is the derivation of the flux ratios of
each order. Instead of relying on pre-determined ratios, based on the
template spectral types, one could derive the flux ratios from the
observed spectra, as done in TODCOR and TRICOR. The best flux ratios
could be derived by maximizing the correlation, using all observed
spectra and therefore can, with high enough SNR, yield better
results. Another improvement, also associated with the flux ratio, is
to apply a ratio that is varying along each order. That is, instead of
the model of Equation (1) for each order, we can use a modified model:

\begin{equation}
R^{(3)}(s_1,s_2,s_3;\alpha(n),\beta(n))=f\ast
\left[ g_1(n-s_1)+\alpha(n)
g_2(n-s_2)+\beta(n) g_3(n-s_3) \right] \ .
\end{equation}
The functions $\alpha(n)$ and $\beta(n)$ could be approximated by
low-degree polynomials. In fact, such an improvement can be applied to
TODCOR, TRICOR and TODMOR as well.

We have applied TRIMOR to the obtained spectra of HD\,188753. Without
investing any additional observational resources we derived the
velocities of the faint component of the triple system and turned the
close pair into double-lined spectroscopic binary. The new radial
velocities led to an estimate of the close pair inclination, which
depends on the total mass of the close pair. We estimated the total
mass of the close pair by three approaches, one that relied on the
spectral type of the two stars of the close pair derived here, one
that used the astrometry of the wide orbit and the mass ratio of the
wide orbit, independent of the present analysis, and finally another
approach that depended on the mass ratio of the close pair derived from
the new velocities of the present analysis and known photometry of the
close pair. All three approaches yielded similar close-pair
inclinations that are close to $41\degr$. The consistency of the
three estimates of the inclination demonstrated the power of TRIMOR
algorithm and established more confidence in the present result.

The derived close-pair inclination suggests that the relative
inclination of HD\,188753 is close to zero and the motion is close to
being coplanar. This result adds HD\,188753 to the increasing set of
triple systems with known relative inclinations \citep[see reviews
by][]{toko04, toko06, toko08}. Applying TRIMOR to more close triple or
quadruple systems \cite[e.g.,][]{shkolnik08}, for which we have at
hand the astrometric orbit of the wide pair from Hipparcos, can
enlarge substantially the set of triples with known relative
inclinations. This certainly will be true after the new GAIA satellite
all-sky astrometry will be available in the near future
\citep[][]{gaia08}.

The distribution of the relative inclinations in triple systems can
shed some light on formation models of binary and triple systems
\citep[e.g.,][]{st02} and can also be relevant to the tidal evolution
of close multiple systems \citep[e.g.,][]{mazeh08}.  Binary and
multiple system formation models of fragmentation
\citep[e.g.,][]{machida08}, which include a later stage of
accretion \citep[e.g.,][]{bbb02, bb06}, predict alignment of the
rotation stellar axes with the orbital angular vector in binaries, as
well as coplanar motion in triple systems. On the other hand, models
assuming that binaries and multiple systems are formed by N-body
interaction \citep[e.g.,][]{clarke96,delgado04} predict that the stellar
rotation might not be aligned with the orbital motion, and triple
systems could have high relative inclination. 

The relative inclination of triple systems might also have implication
for long term evolution of systems for which tidal evolution of the
close pair can circularize the binary orbit during its stellar life
time. \citet{ms79} found that the third star can induce almost
periodic eccentricity modulation into the binary motion, even if the
binary starts with zero eccentricity. (See \citet{kozai62}, who
discovered this effect for asteroids with initial eccentricities.)
\citeauthor{ms79} pointed out that the eccentricity modulation implies
that binaries with close enough distant companions never get into a
stable circular orbit, which means that the binary tidal dissipation
continues to be active forever.  The injected binary eccentricity
invokes frictional forces inside the two stars that dissipate
rotational energy and transfer angular momentum between the stellar
rotation and the binary orbital motion. This causes the semi-major
axis of the binary to shrink on the circularization time scale.

The possible impact of the third star on the shrinking of the close
pair separation depends on the amplitude of the eccentricity
modulation. \citet{ms79} showed that this amplitude does not depend on
the third distant star separation, but strongly depend on the relative
inclination between the two orbits. \citet{eggleton2001},
\cite{eggleton2006}, \cite{bor2004}, and \cite{fabrycky2007} further
studied the effect found by \citet{ms79}, recently termed as the KCTF
(Kozai Cycle with Tidal Friction) effect, and showed that the tidal
effect that leads to the shrinking of the close-pair separation occurs
when the relative angle between the two orbital planes of motion is
larger than $\sim 40^{\circ}$. In fact, \cite{toko06} suggested that
all or at least most of the very close binaries are found with a
distant companion, suggesting that they were formed through the effect
suggested by \citeauthor{ms79} \citep[see, also][for an updated
statistical analysis]{toko08}. It is therefore important to get the
true distribution of relative inclinations of triple systems, with
close pairs in particular. Cases like HD\,188753, with small relative
angle, do not support the shrinking scenario of close binaries in
triple systems, as the eccentricity modulation expected in such cases
is small. As we have shown here, TRIMOR can be very effective in
obtaining the relevant information.

TRIMOR can also be effective in searching spectroscopic evidence for
unknown faint third companions in spectra of known double-lined
binaries. The accumulated data base of multi-order spectra of binaries
can be used for such a project. A somewhat similar approach, with a
different algorithm, was taken by \citet{dan06}, who searched for
spectroscopic trace of a faint tertiary in a large set of observed
spectra of contact binaries. In about a third of the systems they
found an evidence for a tertiary. We note that their spectra are
composed of one order centered on 5200\,\AA, a wavelength for which K
and M stars, for example, are quite faint. In our case, where we had
the luxury of using CORALIE spectra whose orders get up to 6800\,\AA,
we could detect and measure radial velocities of a late K-type
tertiary star. The search for faint companions could be even more
efficient with spectra that go further towards the red wavelegths
\citep[e.g.,][]{fabricant08}.

Finally, TRIMOR can help find extra-solar planets in triple systems,
in the same way TODMOR was instrumental in discovering the planet in
the HD\,41004 system \citep{TODMOR, zucker04}.

\section*{Acknowledgements}
We warmly thank G. Torres for careful reading of the manuscript and
for illuminating suggestions. D. Fabrycky pointed us to the interesting
work of D'Angelo et al.  We thank the referee for very useful comments
that improved substantially the paper.  T.M. acknowledges the
supported by the Israeli Science Foundation (grant no. 655/07).
S.Z. acknowledges financial support by the Israel Science Foundation
(grant No. 119/07). T.M. is indebted to the Radcliffe institute for
advanced studies at Harvard, where the last version of the paper was
prepared. A.E. acknowledges support from the French National Research
Agency through project grant ANR-NT-05-4\_44463.
 
\bibliographystyle{mn2e}
\bibliography{bib_tri82.bib}

\begin{thebibliography}{}

\bibitem[\protect\citeauthoryear{{Baranne}, {Queloz}, {Mayor}, {Adrianzyk},
  {Knispel}, {Kohler}, {Lacroix}, {Meunier}, {Rimbaud} \& {Vin}}{{Baranne}
  et~al.}{1996}]{Baranne96}
{Baranne} A.,  {Queloz} D.,  {Mayor} M.,  {Adrianzyk} G.,  {Knispel} G.,
  {Kohler} D.,  {Lacroix} D.,  {Meunier} J.-P.,  {Rimbaud} G.,    {Vin} A.,
  1996, \aaps, 119, 373

\bibitem[\protect\citeauthoryear{{Bate}, {Bonnell} \& {Bromm}}{{Bate}
  et~al.}{2002}]{bbb02}
{Bate} M.~R.,  {Bonnell} I.~A.,    {Bromm} V.,  2002, \mnras, 336, 705

\bibitem[\protect\citeauthoryear{{Bessell} \& {Brett}}{{Bessell} \&
  {Brett}}{1988}]{bb88}
{Bessell} M.~S.,  {Brett} J.~M.,  1988, \pasp, 100, 1134

\bibitem[\protect\citeauthoryear{{Bonnell} \& {Bate}}{{Bonnell} \&
  {Bate}}{2006}]{bb06}
{Bonnell} I.~A.,  {Bate} M.~R.,  2006, \mnras, 370, 488

\bibitem[\protect\citeauthoryear{{Borkovits}, {Forg{\'a}cs-Dajka} \&
  {Reg{\'a}ly}}{{Borkovits} et~al.}{2004}]{bor2004}
{Borkovits} T.,  {Forg{\'a}cs-Dajka} E.,    {Reg{\'a}ly} Z.,  2004, \aap, 426,
  951

\bibitem[\protect\citeauthoryear{{Carpenter}}{{Carpenter}}{2001}]{carpenter01}
{Carpenter} J.~M.,  2001, \aj, 121, 2851

\bibitem[\protect\citeauthoryear{{Clarke}}{{Clarke}}{1996}]{clarke96}
{Clarke} C.~J.,  1996, \mnras, 283, 353

\bibitem[\protect\citeauthoryear{{D'Angelo}, {van Kerkwijk} \&
  {Rucinski}}{{D'Angelo} et~al.}{2006}]{dan06}
{D'Angelo} C.,  {van Kerkwijk} M.~H.,    {Rucinski} S.~M.,  2006, \aj, 132, 650

\bibitem[\protect\citeauthoryear{{Delgado-Donate}, {Clarke}, {Bate} \&
  {Hodgkin}}{{Delgado-Donate} et~al.}{2004}]{delgado04}
{Delgado-Donate} E.~J.,  {Clarke} C.~J.,  {Bate} M.~R.,    {Hodgkin} S.~T.,
  2004, \mnras, 351, 617

\bibitem[\protect\citeauthoryear{{Eggenberger}, {Udry}, {Mazeh}, {Segal} \&
  {Mayor}}{{Eggenberger} et~al.}{2007}]{Eg07}
{Eggenberger} A.,  {Udry} S.,  {Mazeh} T.,  {Segal} Y.,    {Mayor} M.,  2007,
  \aap, 466, 1179

\bibitem[\protect\citeauthoryear{{Eggleton}}{{Eggleton}}{2006}]{eggleton2006}
{Eggleton} P.,  2006, {Evolutionary Processes in Binary and Multiple Stars}.
Evolutionary Processes in Binary and Multiple Stars, by Peter Eggleton,
  pp.~.~ISBN 0521855578.~Cambridge, UK: Cambridge University Press, 2006.

\bibitem[\protect\citeauthoryear{{Eggleton} \& {Kiseleva-Eggleton}}{{Eggleton}
  \& {Kiseleva-Eggleton}}{2001}]{eggleton2001}
{Eggleton} P.~P.,  {Kiseleva-Eggleton} L.,  2001, \apj, 562, 1012

\bibitem[\protect\citeauthoryear{{Fabricant}, {Kurtz}, {Geller}, {Caldwell},
  {Woods} \& {Dell'Antonio}}{{Fabricant} et~al.}{2008}]{fabricant08}
{Fabricant} D.~G.,  {Kurtz} M.~J.,  {Geller} M.~J.,  {Caldwell} N.,  {Woods}
  D.,    {Dell'Antonio} I.,  2008, \pasp, 120, 1222

\bibitem[\protect\citeauthoryear{{Fabrycky} \& {Tremaine}}{{Fabrycky} \&
  {Tremaine}}{2007}]{fabrycky2007}
{Fabrycky} D.,  {Tremaine} S.,  2007, \apj, 669, 1298

\bibitem[\protect\citeauthoryear{{Griffin}}{{Griffin}}{1967}]{griffin67}
{Griffin} R.~F.,  1967, \apj, 148, 465

\bibitem[\protect\citeauthoryear{{Griffin}}{{Griffin}}{1977}]{Griffin77}
{Griffin} R.~F.,  1977, The Observatory, 97, 15

\bibitem[\protect\citeauthoryear{{Habets} \& {Heintze}}{{Habets} \&
  {Heintze}}{1981}]{hh81}
{Habets} G.~M.~H.~J.,  {Heintze} J.~R.~W.,  1981, \aaps, 46, 193

\bibitem[\protect\citeauthoryear{{Hough}}{{Hough}}{1899}]{Hough99}
{Hough} G.~W.,  1899, Astronomische Nachrichten, 149, 65

\bibitem[\protect\citeauthoryear{{Jang-Condell}}{{Jang-Condell}}{2007}]{jc07}
{Jang-Condell} H.,  2007, \apj, 654, 641

\bibitem[\protect\citeauthoryear{{Jha}, {Torres}, {Stefanik}, {Latham} \&
  {Mazeh}}{{Jha} et~al.}{2000}]{jha00}
{Jha} S.,  {Torres} G.,  {Stefanik} R.~P.,  {Latham} D.~W.,    {Mazeh} T.,
  2000, \mnras, 317, 375

\bibitem[\protect\citeauthoryear{{Konacki}}{{Konacki}}{2005}]{konacki05}
{Konacki} M.,  2005, \nat, 436, 230

\bibitem[\protect\citeauthoryear{{Kozai}}{{Kozai}}{1962}]{kozai62}
{Kozai} Y.,  1962, \aj, 67, 591

\bibitem[\protect\citeauthoryear{{Machida}}{{Machida}}{2008}]{machida08}
{Machida} M.~N.,  2008, \apjl, 682, L1

\bibitem[\protect\citeauthoryear{{Mayor} \& {Mazeh}}{{Mayor} \&
  {Mazeh}}{1987}]{mm87}
{Mayor} M.,  {Mazeh} T.,  1987, \aap, 171, 157

\bibitem[\protect\citeauthoryear{{Mazeh}}{{Mazeh}}{2008}]{mazeh08}
{Mazeh} T.,  2008, in {Goupil} M.-J.,  {Zahn} J.-P.,  eds, EAS Publications
  Series Vol.~29 of EAS Publications Series, {Observational Evidence for Tidal
  Interaction in Close Binary Systems}.
pp 1--65

\bibitem[\protect\citeauthoryear{{Mazeh}, {Martin}, {Goldberg} \&
  {Smith}}{{Mazeh} et~al.}{1997}]{mazeh97}
{Mazeh} T.,  {Martin} E.~L.,  {Goldberg} D.,    {Smith} H.~A.,  1997, \mnras,
  284, 341

\bibitem[\protect\citeauthoryear{{Mazeh} \& {Shaham}}{{Mazeh} \&
  {Shaham}}{1979}]{ms79}
{Mazeh} T.,  {Shaham} J.,  1979, \aap, 77, 145

\bibitem[\protect\citeauthoryear{{Mazeh}, {Simon}, {Prato}, {Markus} \&
  {Zucker}}{{Mazeh} et~al.}{2003}]{mazeh03}
{Mazeh} T.,  {Simon} M.,  {Prato} L.,  {Markus} B.,    {Zucker} S.,  2003,
  \apj, 599, 1344

\bibitem[\protect\citeauthoryear{{Mazeh}, {Zucker}, {Goldberg}, {Latham},
  {Stefanik} \& {Carney}}{{Mazeh} et~al.}{1995}]{mazeh95}
{Mazeh} T.,  {Zucker} S.,  {Goldberg} D.,  {Latham} D.~W.,  {Stefanik} R.~P.,
   {Carney} B.~W.,  1995, \apj, 449, 909

\bibitem[\protect\citeauthoryear{{Metcalfe}, {Mathieu}, {Latham} \&
  {Torres}}{{Metcalfe} et~al.}{1996}]{metcalfe96}
{Metcalfe} T.~S.,  {Mathieu} R.~D.,  {Latham} D.~W.,    {Torres} G.,  1996,
  \apj, 456, 356

\bibitem[\protect\citeauthoryear{{Pepe}, {Mayor}, {Delabre}, {Kohler},
  {Lacroix}, {Queloz}, {Udry}, {Benz}, {Bertaux} \& {Sivan}}{{Pepe}
  et~al.}{2000}]{pepe00}
{Pepe} F.,  {Mayor} M.,  {Delabre} B.,  {Kohler} D.,  {Lacroix} D.,  {Queloz}
  D.,  {Udry} S.,  {Benz} W.,  {Bertaux} J.-L.,    {Sivan} J.-P.,  2000, in
  {Iye} M.,  {Moorwood} A.~F.,  eds, Proc. SPIE Vol. 4008, p. 582-592, Optical
  and IR Telescope Instrumentation and Detectors, Masanori Iye; Alan F.
  Moorwood; Eds. Vol.~4008 of Presented at the Society of Photo-Optical
  Instrumentation Engineers (SPIE) Conference, {HARPS: a new high-resolution
  spectrograph for the search of extrasolar planets}.
pp 582--592

\bibitem[\protect\citeauthoryear{{Perryman}, {de Bruijne} \&
  {Lammers}}{{Perryman} et~al.}{2008}]{gaia08}
{Perryman} M.,  {de Bruijne} J.,    {Lammers} U.,  2008, Experimental
  Astronomy, 22, 143

\bibitem[\protect\citeauthoryear{{Pfahl}}{{Pfahl}}{2005}]{pfahl05}
{Pfahl} E.,  2005, \apjl, 635, L89

\bibitem[\protect\citeauthoryear{{Pickles}}{{Pickles}}{1998}]{Pickles98}
{Pickles} A.~J.,  1998, \pasp, 110, 863

\bibitem[\protect\citeauthoryear{{Pont}, {Zucker} \& {Queloz}}{{Pont}
  et~al.}{2006}]{pont06}
{Pont} F.,  {Zucker} S.,    {Queloz} D.,  2006, \mnras, 373, 231

\bibitem[\protect\citeauthoryear{{Prato}, {Simon}, {Mazeh}, {McLean}, {Norman}
  \& {Zucker}}{{Prato} et~al.}{2002}]{prato02}
{Prato} L.,  {Simon} M.,  {Mazeh} T.,  {McLean} I.~S.,  {Norman} D.,
  {Zucker} S.,  2002, \apj, 569, 863

\bibitem[\protect\citeauthoryear{{Scargle}}{{Scargle}}{1982}]{scargle82}
{Scargle} J.~D.,  1982, \apj, 263, 835

\bibitem[\protect\citeauthoryear{{Shkolnik}, {Liu}, {Reid}, {Hebb}, {Cameron},
  {Torres} \& {Wilson}}{{Shkolnik} et~al.}{2008}]{shkolnik08}
{Shkolnik} E.,  {Liu} M.~C.,  {Reid} I.~N.,  {Hebb} L.,  {Cameron} A.~C.,
  {Torres} C.~A.,    {Wilson} D.~M.,  2008, \apj, 682, 1248

\bibitem[\protect\citeauthoryear{{Shporer}, {Hartman}, {Mazeh} \&
  {Pietsch}}{{Shporer} et~al.}{2007}]{shporer07}
{Shporer} A.,  {Hartman} J.,  {Mazeh} T.,    {Pietsch} W.,  2007, \aap, 462,
  1091

\bibitem[\protect\citeauthoryear{{Simkin}}{{Simkin}}{1974}]{simkin74}
{Simkin} S.~M.,  1974, \aap, 31, 129

\bibitem[\protect\citeauthoryear{{Smith} \& {Gray}}{{Smith} \&
  {Gray}}{1976}]{sg76}
{Smith} M.~A.,  {Gray} D.~F.,  1976, \pasp, 88, 809

\bibitem[\protect\citeauthoryear{{S{\"o}derhjelm}}{{S{\"o}derhjelm}}{1999}]{so%
derhjelm99}
{S{\"o}derhjelm} S.,  1999, \aap, 341, 121

\bibitem[\protect\citeauthoryear{{Sterzik} \& {Tokovinin}}{{Sterzik} \&
  {Tokovinin}}{2002}]{st02}
{Sterzik} M.~F.,  {Tokovinin} A.~A.,  2002, \aap, 384, 1030

\bibitem[\protect\citeauthoryear{{Tokovinin}}{{Tokovinin}}{2004}]{toko04}
{Tokovinin} A.,  2004, in {Allen} C.,  {Scarfe} C.,  eds, Revista Mexicana de
  Astronomia y Astrofisica Conference Series Vol.~21 of Revista Mexicana de
  Astronomia y Astrofisica, vol. 27, {Statistics of multiple stars}.
pp 7--14

\bibitem[\protect\citeauthoryear{{Tokovinin}}{{Tokovinin}}{2008}]{toko08}
{Tokovinin} A.,  2008, \mnras, 389, 925

\bibitem[\protect\citeauthoryear{{Tokovinin}, {Thomas}, {Sterzik} \&
  {Udry}}{{Tokovinin} et~al.}{2006}]{toko06}
{Tokovinin} A.,  {Thomas} S.,  {Sterzik} M.,    {Udry} S.,  2006, \aap, 450,
  681

\bibitem[\protect\citeauthoryear{{Tokovinin} \& {Gorynya}}{{Tokovinin} \&
  {Gorynya}}{2007}]{tokovinin07}
{Tokovinin} A.~A.,  {Gorynya} N.~A.,  2007, \aap, 465, 257

\bibitem[\protect\citeauthoryear{{Tonry} \& {Davis}}{{Tonry} \&
  {Davis}}{1979}]{td79}
{Tonry} J.,  {Davis} M.,  1979, \aj, 84, 1511

\bibitem[\protect\citeauthoryear{{Torres}, {Lacy}, {Marschall}, {Sheets} \&
  {Mader}}{{Torres} et~al.}{2006}]{torres06}
{Torres} G.,  {Lacy} C.~H.,  {Marschall} L.~A.,  {Sheets} H.~A.,    {Mader}
  J.~A.,  2006, \apj, 640, 1018

\bibitem[\protect\citeauthoryear{{Torres}, {Latham} \& {Stefanik}}{{Torres}
  et~al.}{2007}]{torres07}
{Torres} G.,  {Latham} D.~W.,    {Stefanik} R.~P.,  2007, \apj, 662, 602

\bibitem[\protect\citeauthoryear{{Torres}, {Stefanik} \& {Latham}}{{Torres}
  et~al.}{1997}]{torres97}
{Torres} G.,  {Stefanik} R.~P.,    {Latham} D.~W.,  1997, \apj, 479, 268

\bibitem[\protect\citeauthoryear{{Torres}, {Stefanik}, {Latham} \&
  {Mazeh}}{{Torres} et~al.}{1995}]{torres95}
{Torres} G.,  {Stefanik} R.~P.,  {Latham} D.~W.,    {Mazeh} T.,  1995, \apj,
  452, 870

\bibitem[\protect\citeauthoryear{{Zechmeister} \& {K{\"u}rster}}{{Zechmeister}
  \& {K{\"u}rster}}{2009}]{kurster09}
{Zechmeister} M.,  {K{\"u}rster} M.,  2009, \aap, 496, 577

\bibitem[\protect\citeauthoryear{{Zucker}}{{Zucker}}{2003}]{Zucker03}
{Zucker} S.,  2003, \mnras, 342, 1291

\bibitem[\protect\citeauthoryear{{Zucker}}{{Zucker}}{2004}]{zucker2004}
{Zucker} S.,  2004, in {Hilditch} R.~W.,  {Hensberge} H.,   {Pavlovski} K.,
  eds, Spectroscopically and Spatially Resolving the Components of the Close
  Binary Stars Vol.~318 of Astronomical Society of the Pacific Conference
  Series, {TODCOR - TwO-Dimensional CORrelation}.
pp 77--85

\bibitem[\protect\citeauthoryear{{Zucker} \& {Mazeh}}{{Zucker} \&
  {Mazeh}}{1994}]{zm94}
{Zucker} S.,  {Mazeh} T.,  1994, \apj, 420, 806

\bibitem[\protect\citeauthoryear{{Zucker}, {Mazeh}, {Santos}, {Udry} \&
  {Mayor}}{{Zucker} et~al.}{2003}]{TODMOR}
{Zucker} S.,  {Mazeh} T.,  {Santos} N.~C.,  {Udry} S.,    {Mayor} M.,  2003,
  \aap, 404, 775

\bibitem[\protect\citeauthoryear{{Zucker}, {Mazeh}, {Santos}, {Udry} \&
  {Mayor}}{{Zucker} et~al.}{2004}]{zucker04}
{Zucker} S.,  {Mazeh} T.,  {Santos} N.~C.,  {Udry} S.,    {Mayor} M.,  2004,
  \aap, 426, 695

\bibitem[\protect\citeauthoryear{{Zucker}, {Torres} \& {Mazeh}}{{Zucker}
  et~al.}{1995}]{TRICOR}
{Zucker} S.,  {Torres} G.,    {Mazeh} T.,  1995, \apj, 452, 863

\end{thebibliography}
 
\label{lastpage}

\end{document}